\documentclass{article}
\usepackage{spconf,amsmath,graphicx}
\usepackage{pdfpages}
\usepackage{hyperref}
\usepackage[numbers,sort,compress]{natbib} % sort citations

\def\prepara{{\vspace{5pt}}}

\title{VoxSRC 2019: The first VoxCeleb Speaker Recognition Challenge}

\name{\em Joon Son Chung$^{1,2}$, Arsha Nagrani$^1$, Ernesto Coto$^1$, Weidi Xie$^1$, Mitchell McLaren$^3$, \\ 
\em Douglas A Reynolds$^4$ and Andrew Zisserman$^1$}

\address{$^1$Visual Geometry Group, Department of Engineering Science, University of Oxford, UK\\
$^2$Naver Corporation, South Korea \\
$^3$Speech Technology and Research Laboratory, SRI International, Menlo Park, CA, USA \\
$^4$MIT Lincoln Laboratory, Lexington, MA, USA \\
\small{\url{http://www.robots.ox.ac.uk/\~vgg/data/voxceleb/competition.html}}}

\begin{document}
\maketitle
\begin{abstract}
The VoxCeleb Speaker Recognition Challenge 2019 aimed to assess how well current speaker recognition technology is able to identify speakers in unconstrained or `in the wild' data. It consisted of:  (i) a publicly available speaker recognition dataset from YouTube videos together with ground truth annotation and standardised evaluation  software;  and  (ii)  a  public  challenge  and  workshop held at Interspeech 2019 in Graz, Austria.
This paper outlines the challenge and provides its baselines, results and discussions. 
\end{abstract}

\section{Introduction}
\label{sec:intro}

The VoxCeleb Speaker Recognition Challenge (VoxSRC) 2019 was the first of a new series of speaker recognition challenges that are intended to be hosted annually. VoxSRC 2019 consisted of: (i) a publicly available speaker recognition dataset with speech segments `in the wild', together with ground truth annotations and standardised evaluation software; and (ii) a public challenge and workshop held at Interspeech 2019 in Graz, Austria. The VoxSRC challenge series is intended to: (i) explore and promote new research in speaker recognition ‘in the wild’; (ii) measure and calibrate the performance of the current state of technology through public evaluation tools; and (iii) provide an open-source free dataset accessible to all. 

While speech technologies have developed rapidly during the last few decades (with a large focus on speaker verification), speaker recognition under noisy and unconstrained conditions
is still an extremely challenging topic. Applications of speaker
recognition are many and varied, ranging from authentication
in high-security systems and forensic tests, to high fidelity search of persons in large corpora of speech data. For such systems to be deployed in the real world, it is crucial that they work under unconstrained conditions, with noisy, varied and sometimes very short and fleeting speech segments. The 1st VoxCeleb Speaker Recognition Challenge (VoxSRC 2019) aims to assess how well current speech verification technology is able to identify similar speakers across those challenging conditions and to explore novel approaches on this task with fixed experimental conditions and training data.

The VoxSRC challenge is heavily inspired by the Speakers In the Wild (SITW) challenge~\cite{mclaren2015speakers}, and is complementary to other challenge series such as those of National Institute in Standards of Technology (NIST)~\cite{alvin2004nist} and ASVspoof~\cite{wu2017asvspoof}. 
There are two main differences between VoxSRC and the latest NIST Speaker
Recognition Evaluations~\cite{alvin2004nist} (SRE): (i) For the VoXSRC \textit{fixed} condition (described in detail in Sec.~\ref{sec:tracks}), there was no explicitly induced domain-shift between development and evaluation data; and (ii) the audio segments involved in the pair-wise speaker comparisons (i.e., trials) in VoxSRC have a much shorter average duration, making the task more challenging. Additionally, all the training and validation data was (and will continue to be) free and available to researchers irrespective of whether they entered the challenge or not. The workshop was also free for participants to attend. 

In this paper, we describe the details of the evaluation task, the datasets provided, the challenge evaluation results and subsequent discussion. Further details can be found at the challenge website \footnote{\url{http://www.robots.ox.ac.uk/~vgg/data/voxceleb/competition.html}}. 
% We also append the technical reports of the challenge winners to this document. \\
% \textbf{How to cite this paper:} If you would like to cite the challenge as a whole, or are submitting a paper referencing your performance in the challenge, please use this paper as a reference. However if you are referencing any of the methods submitted to the challenge, please check whether the original authors of the systems have uploaded a paper describing their method, and give them the appropriate credit.

\section{Task Description} 
The task is \textit{speaker verification}, where given pairs of audio segments, the goal is to simply say whether they are from the same speaker or from different speakers.
\subsection{Tracks} \label{sec:tracks}
The challenge consisted of the following two tracks:
\begin{enumerate}
    \item Speaker Verification -- fixed training condition
    \item Speaker Verification -- open training condition
\end{enumerate}
The fixed training condition required that participants train only on the VoxCeleb2 dev dataset~\cite{Chung18a}, which contains 1,092,009 utterances from 5,994 speakers. For the open training condition, participants could use the VoxCeleb datasets and any other data (including that which is not publicly released) except for the challenge's \textit{test} data. The data is described in more detail in the next section. 

\subsection{Data}
The VoxCeleb datasets~\cite{nagrani2020voxceleb,Chung18a,Nagrani17} consist of speech segments from unconstrained YouTube videos for several thousand individuals. These videos contain clean studio interviews, red carpet interviews, outdoor and noisy conditions and multi-speaker scenarios. While some videos are professionally recorded, others are acquired using hand-held or other crude recording devices with no editing. The VoxCeleb datasets were acquired using an automatic pipeline based on computer vision techniques. For a full description of the pipeline and an overview of the datasets, see~\cite{nagrani2020voxceleb}. All noise, reverberation, compression and other artifacts in the corpus are natural characteristics of the original audio and have not been removed. Since the speech is conversational, the segments are short and quick and may suffer from some background speech from other identities. The duration of each speech segment is unconstrained, as is the total amount of speech present per speaker. We believe that recognizing the same speaker across such varied conditions is representative of many challenges that speaker verification technology will have to face today if it is to be deployed in the real world. 

\prepara\noindent\textbf{Training data:} The fixed training condition required that participants train only on the VoxCeleb2 dev dataset~\cite{Chung18a}, which contains 1,092,009 utterances from 5,994 speakers. For the open training condition, participants could use the VoxCeleb datasets and any other data except for the challenge's \textit{test} data. In order to encourage participation from industry, and to benchmark absolute best performance, there were no restrictions on the data that could have been used for the open condition, in particular the data did not have to be made publicly available to the research community. 

\prepara\noindent\textbf{Validation data:} We encouraged participants to validate their models using the VoxCeleb1 publicly released hard and easy test lists: VoxCeleb1, VoxCeleb1-E, and VoxCeleb1-H. A ‘cleaned’ version of each was also made available in which the data was checked manually for any errors using the same procedure described in \cite{nagrani2020voxceleb} (and briefly described below).
% \begin{enumerate}
%     \item VoxCeleb1
%     \item VoxCeleb1 (cleaned)
%     \item VoxCeleb1-H
%     \item VoxCeleb1-H (cleaned)
%     \item VoxCeleb1-E
%     \item VoxCeleb1-E (cleaned)
% \end{enumerate}
These lists can be found at this url: \url{http://www.robots.ox.ac.uk/~vgg/data/voxceleb/vox2.html}. 

\prepara\noindent\textbf{Test data:} The test data was created from YouTube videos in the same way as the training and validation sets, however unlike the training data, the test data was subjected to an additional manual verification step. We released the test data a month before the challenge results were due, but it was \textit{blind}, i.e.\ the speech segments were released but with no annotations. The test data was released strictly for reporting of results alone, participants were not allowed to use this data in any way to train or tune systems. The statistics of the test set can be found in Table \ref{tab:testdata}. 
The test data was checked manually for any errors using the same procedure described in \cite{nagrani2020voxceleb}. This was done using a simple web-based tool that shows all video segments for each identity. In order to highlight the segments which are more likely to contain errors,
face and voice embeddings were generated from SphereFace~\cite{liu2017sphereface} and our own model trained on {\tt VoxCeleb2} respectively,
and those with lower confidence were highlighted with a different colour to aid manual inspection.

Since the test pairs were sampled at random, the VoxCeleb test sets include a proportion (approx.\ 10\%) of same-session trials; that is, a comparison of audio samples extracted from different parts of the same original video. Such trials are easier than those from different sessions, due to the lack of change in intrinsic speaker traits that typically occur over time. Future instances of the challenge will not have same-session trials in the test and validation sets.  

\begin{table}[]
    \centering
    
    \begin{tabular}{c|c|c|c}
         \textbf{\# Speakers} & \textbf{\# Pairs} & \textbf{\# Utter.} & \textbf{Segment length (s)}  \\ \hline
         745 & 208,008 & 19,154 & 3.92/ 7.42/ 81.04
    \end{tabular}
    \caption{\small{ Statistics of the test set. \textbf{\# Pairs} refers to the number of evaluation trial pairs, whereas \textbf{\# Utter.} refers to the total number of unique speech segments in the test set. Segment lengths are reported as min/mean/max.}}
    \label{tab:testdata}
\end{table}

\begin{table*}[]
    \centering
    
    \begin{tabular}{c|c|c|c}
         \textbf{Rank} & \textbf{Team Name} & \textbf{Organization} & \textbf{EER}  \\ \hline
         1 & BUT~\cite{BUT} & Brno Univ. of Technology, Speech@FIT and IT4I Center of Excellence, Brno & 0.0142\\ 
         2 & JHU-HLTCOE~\cite{JHU} & Human Language Technology Center of Excellence, Johns Hopkins Univ., Baltimore & 0.0154 \\ 
         3 & TZ~\cite{Microsoft} & Microsoft Corp. Redmond, USA & 0.0171 \\ 
    \end{tabular}
    \caption{\small{Winners for track 1 (fixed training condition). References refer to winners' presentation slides. All slides can be found at the challenge website.}}
    \label{tab:results_fixed}
\end{table*}

\begin{table*}[]
    \centering
    
    \begin{tabular}{c|c|c|c}
         \textbf{Rank} & \textbf{Team Name} & \textbf{Organization} & \textbf{EER}  \\ \hline
         1 & BUT~\cite{BUT} & Brno Univ. of Technology, Speech@FIT and IT4I Center of Excellence, Brno & 0.0126\\ 
         2 & TZ~\cite{Microsoft} & Microsoft Corp. Redmond, USA & 0.0149 \\ 
         3 & JHU-HLTCOE~\cite{JHU} & Human Language Techn. Center of Excellence, Johns Hopkins Univ., Baltimore & 0.0169 \\
         3 & DKU-TVM-SYSU~\cite{DKU} & Duke Kunshan Univ., Sun Yat-sen Univ., TV Mining Media Tech. Ltd, China & 0.0169 \\
    \end{tabular}
    \caption{\small{Winners for track 2 (open training condition). References refer to winners' presentation slides. All slides can be found at the challenge website.}}
    \label{tab:results_open}
\end{table*}

\section{Challenge Mechanics} 
\subsection{Evaluation and scoring}
The evaluation data protocol comprised of a list of trials, each corresponding to
a pair of audio segments. Participants were asked to assign to
each trial a real-valued, finite, floating-point similarity score. All scores were required to be in the closed interval [0, 1], where a similarity score of 1 means the pair of segments correspond to the same speaker and 0 means the pair of segments correspond to different speakers. System output was evaluated using the Equal Error Rate (EER). 
\subsubsection{Metric - Equal Error Rate}
For a single evaluation pair $x_1$ and $x_2$, the miss (or false rejection FR) probability is given by 
$$P_{FR}(\theta) = P(s<\theta|y_1 = y_2)$$
where $\theta$ is the accept/reject decision threshold and $s$ is the similarity score for the hypothesis that the speaker of the first test utterance $y_1$ is equal to the second one $y_2$. For a given decision threshold $\theta$, the false acceptance (FA) probability is given by 
$$P_{FA}(\theta) = P(s>\theta|y_1 \neq y_2)$$
Acceptance (of the two segments as belonging to the same identity) is made if the score $s$ is above the threshold $\theta$ and rejection occurs when the score $s$ is below the threshold $\theta$.
Equal error rate (EER) is used to determine the threshold value for a system when its false acceptance rate (FAR) and false rejection rate (FRR) are equal. This rate is then known as the Equal Error Rate.

Participants were only allowed one submission per day, and only three submissions in total in order to prevent overfitting to the test set.

\subsection{Baseline} 
We provide a simple CNN based baseline, which is trained on input spectrograms. The baseline is described in detail in~\cite{Nagrani17}, and is trained on the VoxCeleb2 dev set. The baseline achieved obtained an EER of 0.1140 on the test set.   

\subsection{Challenge and Permanent Phase} 
The challenge server was hosted using CodaLab\footnote{\url{https://competitions.codalab.org/}}. Public leaderboards are displayed online for two phases: (i) the challenge phase which ended on Aug. 30, 2019, and includes submissions to be taken into account for the challenge; and (ii) the permanent phase, which allows participants to measure their performance on the test set even after the challenge is over. Indeed, at the time of writing, there have been 11 new submissions for the open training condition and 20 new submissions for the fixed training condition since the end of the challenge phase. Once again, in order to prevent overfitting of the test set, participants are only allowed one submission per day and up to 10 submissions in total for the permanent phase.
\section{Methods and Results} 
Although this was the first year of the challenge, over 50 teams from 17 countries participated in total, with 50 teams submitting to the fixed track, and 35 to the open track. The winners for both tracks are shown in Tables \ref{tab:results_fixed} and \ref{tab:results_open}. 
%and we append the technical reports of the winners to this document.  
We were pleasantly surprised to see that 90\% of submissions beat our baseline provided in the fixed condition, and 85\% beat the baseline in the open condition (indeed our baseline performed quite poorly in the challenge). 
Please refer to the challenge website for the full leaderboard of results. 

\begin{figure}[t]
\begin{center}
\includegraphics[width=0.45\textwidth]{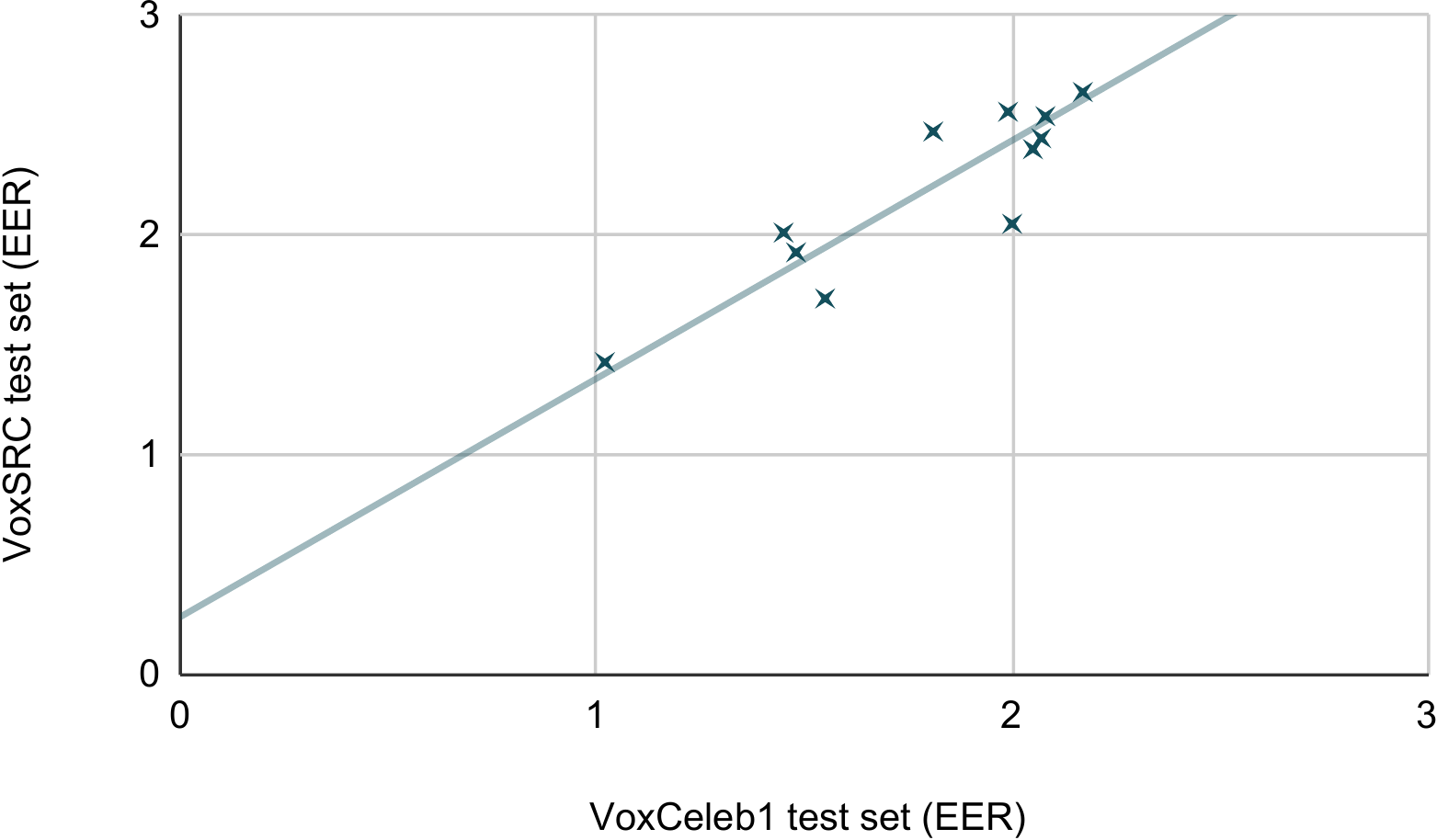}
\end{center}
   \caption{\small{Correlation between validation performance on the VoxCeleb1 test set and final performance on the VoxSRC hidden test set. We plot the performance of 11 different systems, as reported in the technical reports submitted to the challenge,}}
\label{fig:cor}
\end{figure}

The winning methods used deep neural networks (DNN) based systems with a division into front end and back end components. The front end system consisted of an embedding extraction network, which maps a variable length speech segment into a fixed length embedding. The back end system then consisted of a classification head and a scoring procedure. 
A number of different DNNs were then used as feature extractors, including variations on 1-D convolution based TDNNs~\cite{waibel1989phoneme} (variations including different sizes and adding residual connections) or ResNets~\cite{he2016deep} with 2D convolutions. In both cases the inputs used for the front end systems were MFCCs or alternate acoustic features such as short-term 2D spectrograms (as opposed to raw audio signals). Winning submissions found that it was easy to overfit on the training set, and hence used techniques like heavy augmentation of the training data, in particular augmentation stratgies provided by the Kaldi recipe\footnote{\url{https://github.com/kaldi-asr/kaldi/tree/master/egs/sre16/v2}}, (i.e.\ Room Impulse Response (RIR) and the Musan~\cite{snyder2015musan} noise dataset), and network regularization. Margin penalties were also found to be an effective training strategy, with popular losses of choice being the additive angular margin (AAM) loss~\cite{deng2019arcface} and the additive margin softmax loss \cite{wang2018additive}.
Utterance level aggregation of the embedding vectors used both the mean and variance of the features, and popular back end scoring systems were G-PLDA~\cite{garcia2012multicondition} or Cosine distance scoring, with adaptive symmetric score normalization. Almost all the top submissions fused multiple networks in an ensemble by (weighted) averaging. Interesting findings are that: TDNN and ResNets are complementary~\cite{BUT}; and that
fusing smaller heterogeneous systems outperforms a large DNN
with a similar number of parameters~\cite{JHU}.
Other significant ideas were phonetic attention that can deemphasize
the importance of frames depending on the speech content, e.g.\ to
weight non-speech segments lower~\cite{Microsoft}; although this had
limited impact on VoxSRC as almost all the utterances contain little silence. For the same reason, speech activity detection
(SAD) also had little impact~\cite{JHU}.
Interestingly, the best performance in the open condition (EER = 0.0126) was only marginally superior to that achieved in the fixed condition (EER = 0.0142), suggesting that additional training data may not be useful due to domain transfer issues. Additional training data used by winning teams included the development part of VoxCeleb-1 (1,152 speakers), 2338 speakers from the LibriSpeech dataset~\cite{panayotov2015librispeech} and  1735 speakers from the Deep-Mine dataset~\cite{zeinali2018deepmine}. 
We also note that many participants used the VoxCeleb1 test set as a validation set to caliberate system performance. An analysis of the correlation between performance on the VoxCeleb1 test set and the hidden VoxSRC test set can be found in Fig. \ref{fig:cor}. Given 11 data points, we find that performance on the validation set is a good indication of final hidden test set performance.

\section{Workshop}
The VoxSRC 2019 workshop was held in Graz, Austria, in conjunction with Interspeech 2019, and was well attended with over 100 attendees. Presentations at the workshop consisted of a keynote by Mitchell McLaren summarising the history of speaker verification methods and benchmarks, as well as short talks by the winners of both tracks summarising their methods. The workshop concluded with a productive discussion on ways to improve future versions of VoxSRC, which are summarised in Section~\ref{sec:discussion}. All of the presentation slides are available at \url{http://www.robots.ox.ac.uk/~vgg/data/voxceleb/interspeech2019.html}. The workshop was kindly sponsored by Naver Corpororation and the Oxford Wave Institute. 

\section{Discussion} \label{sec:discussion}
In this section, we discuss some limitations of the first VoxSRC challenge and future plans. First, we used only a single metric -- EER to evaluate performance. Further challenges will incorporate other metrics such as the minimum normalized Detection Cost (DCF) at two operating points with $P_{Target}=10^{-3}$ (DCF1) and $P_{Target}=10^{-2}$ (DCF2). This is to allow the \textit{calibration} of a system to impact performance, as calibration is a vital aspect to many real-world deployments of speaker verification technology.
Second, given that this was the first instance of the challenge, we focused on having only a single task -- audio only speaker verification. Given recent academic interest in audio-visual multimodal and cross-modal person recognition~\cite{sell2018audio,giraudel2012repere,Nagrani18a}, in future versions of the challenge we plan to include an audio-visual task wherein the faces of speakers will also be involved in the challenge. Additional audio-only tasks that can be incorporated in future challenges include speaker detection (temporally locating a single speaker from a multispeaker speech segment) and speaker diarisation (splitting up a multispeaker speech segment into temporal boundaries based on identity, i.e.\ solving who speaks `when'). Both these additional tasks, however, require expensive manual annotation of speech segments.
We also note that after the first challenge, due to excellent submissions from over 50 teams, performance on our test set is almost saturated (EER = 0.0126), and hence in future challenges we will endeavour to have a more challenging test set. 

\subsection*{Acknowledgements}
This work is funded by the EPSRC Programme
Grant Seebibyte EP/M013774/1. Arsha is funded by a Google PhD Fellowship. This material is based upon work supported by the Air Force Research Laboratory under Air Force Contract No. FA8702-15-D-0001.  Any opinions, findings and conclusions or recommendations expressed in this material are those of the author(s) and do not necessarily reflect the views of the US Department of Defense. 
We would like to thank Daniel Garcia-Romero for his comments, and also Triantafyllos Afouras, Hanna Ryu, Hansol Kim and Young Jae Kim for their help with the logistics and organisation of the workshop.

\bibliographystyle{IEEEbib}
\bibliography{shortstrings,refs}

% \clearpage
% \includepdf[pages=-]{FIXED_1st.pdf}
% \includepdf[pages=-]{FIXED_2nd.pdf}
% \includepdf[pages=-]{FIXED_3rd.pdf}
% \includepdf[pages=-]{OPEN_1st.pdf}
% \includepdf[pages=-]{OPEN_2nd.pdf}
% \includepdf[pages=-]{OPEN_3rdEQUAL1.pdf}
% \includepdf[pages=-]{OPEN_3rdEQUAL2.pdf}

\end{document}